\documentclass[twocolumn, aps, superscriptaddress]{revtex4}

\usepackage{graphicx}
\usepackage{amsmath,amssymb}
\usepackage{epstopdf}

\usepackage{tikz}
\usetikzlibrary{shapes,arrows}

\newcommand{\beq}{\begin{equation}}
\newcommand{\eeq}{\end{equation}}
\newcommand{\bdm}{\begin{displaymath}}
\newcommand{\edm}{\end{displaymath}}

\DeclareFontFamily{OT1}{pzc}{}
\DeclareFontShape{OT1}{pzc}{m}{it}{<-> s * [1.10] pzcmi7t}{}
\DeclareMathAlphabet{\mathpzc}{OT1}{pzc}{m}{it}

\graphicspath{{./plots/}}
\begin{document}

\title{Limiting the effects of earthquakes on gravitational-wave interferometers}

\author{Michael Coughlin}
\affiliation{Department of Physics, Harvard University, Cambridge, MA 02138, USA}

\author{Paul Earle}
\affiliation{U.S. Geological Survey, Golden, CO 80401, USA}

\author{Jan Harms}
\affiliation{INFN, Sezione di Firenze, Sesto Fiorentino, 50019, Italy\\
Universit\`a degli Studi di Urbino ``Carlo Bo'', I-61029 Urbino, Italy}

\author{Sebastien Biscans}
\affiliation{LIGO Laboratory, Massachusetts Institute of Technology, Cambridge, MA 02138, USA}

\author{Christopher Buchanan}
\affiliation{Department of Physics and Astronomy, Louisiana State University, Baton Rouge, LA 70803-4001, USA}

\author{Eric Coughlin}
\affiliation{Department of Computer Science, Luther College, 700 College Dr, Decorah, IA 52101, USA}

\author{Fred Donovan}
\affiliation{LIGO Laboratory, Massachusetts Institute of Technology, Cambridge, MA 02138, USA}

\author{Jeremy Fee}
\affiliation{United States Geological Survey, Golden, CO 80401, USA}

\author{Hunter Gabbard}
\affiliation{Albert-Einstein-Institut, Max-Planck-Institut f{\"u}r Gravitationsphysik, D-30167 Hannover, Germany}

\author{Michelle Guy}
\affiliation{United States Geological Survey, Golden, CO 80401, USA}

\author{Nikhil Mukund}
\affiliation{Inter-University Centre for Astronomy and Astrophysics (IUCAA), Post Bag 4, Ganeshkhind,  Pune 411 007, India}

\author{Matthew Perry}
\affiliation{Planetary Science Institute, Lakewood, CO 80401, USA}

\begin{abstract}

Ground-based gravitational wave interferometers such as the Laser Interferometer Gravitational-wave Observatory (LIGO) are susceptible to ground shaking from high-magnitude teleseismic events, which can interrupt their operation in science mode and significantly reduce their duty cycle. It can take several hours for a detector to stabilize enough to return to its nominal state for scientific observations. The down time can be reduced if advance warning of impending shaking is received and the impact is suppressed in the isolation system with the goal of maintaining stable operation even at the expense of increased instrumental noise. Here, we describe an early warning system for modern gravitational-wave observatories. The system relies on near real-time earthquake alerts provided by the U.S. Geological Survey (USGS) and the National Oceanic and Atmospheric Administration (NOAA).  Preliminary low latency hypocenter and magnitude information is generally available in 5 to 20 minutes of a significant earthquake depending on its magnitude and location.  The alerts are used to estimate arrival times and ground velocities at the gravitational-wave detectors.  In general, 90\% of the predictions for ground-motion amplitude are within a factor of 5 of measured values.  The error in both arrival time and ground-motion prediction introduced by using preliminary, rather than final, hypocenter and magnitude information is minimal. By using a machine learning algorithm, we develop a prediction model that calculates the probability that a given earthquake will prevent a detector from taking data.   Our initial results indicate that by using detector control configuration changes, we could prevent interruption of operation from 40 to 100 earthquake events in a 6-month time-period.

\end{abstract}

\maketitle

\section{Introduction}
\label{sec:Intro}

Earthquakes are a significant issue for gravitational-wave detectors. In previous work \cite{CoSt2015}, Coughlin et al. described how large-scale astronomical experiments, such as meter class telescopes and gravitational-wave interferometers, are susceptible to earthquakes. In the case of telescopes, the predominant concern is the potential for nearby, devastating earthquakes, which will damage either the surrounding structure or the mirrors that make up the telescope, and it is argued that a regional early earthquake warning (EEW) \cite{Al2012,KuAl2013a,KuAl2013b,KuHe2014,CoLa2009a,CoLa2009b,BoAl2014,HoKa2008,HoEA2011c,StAl2016} system is important to minimize potential damage to telescopes. 
Gravitational-wave detectors, on the other hand, are susceptible to teleseismic events from around the world \cite{MaFa2012}. 
The two detectors of the Laser Interferometer Gravitational-wave Observatory (LIGO) \cite{aligo} that have made the first direct observations of gravitational waves \cite{AbEA2016a,AbEA2016e} form a global network of gravitational-wave interferometers together with the Virgo \cite{avirgo}, and GEO600 \cite{Gr2010} detectors. These detectors can be destabilized by significant ground motion, despite seismic isolation systems designed to minimize such effects \cite{AbAd2002,StAb2009,MaLa2015}.

During the last LIGO science run, large amplitude earthquakes from around the world would typically cause the detectors to fall out of lock \cite{CoSt2015}, which signifies a failure of the control system to maintain optics at their nominal positions and orientations with subsequent loss of laser power in the system. Not only were the data around the time of the earthquakes not useful for gravitational-wave detection, but it would also take hours of dead time for the detectors to return to the locked state. 
We showed that there are potential gains to be made with an early warning system assuming that the incurred downtime could be reduced with sufficient advance notice of the earthquakes' arrivals.
Detailed studies of earthquake response during previous science runs showed that there is about one teleseismic event each week producing ground motion at the sites too strong for the control system to be able to maintain lock. In most cases, it was then impossible to lock the interferometer for some hours. A scheme that would suppress disturbances of earthquakes early in the isolation system with the final goal to maintain lock during strong ground motion, even at the price of increased instrumental noise, could potentially lead to substantial increase of the duty cycle. This will likely be of greater importance even in high-power configurations of the advanced detectors, where thermalization of test masses during the locking procedure could potentially increase the time it takes to reach maximal sensitivity.

For this reason, we have created an earthquake early warning client named \emph{Seismon}, which uses a real-time event messaging system of the U.S. Geological Survey (USGS) to mitigate the effects of teleseismic events on ground-based gravitational-wave detectors. The messages contain information about the earthquake source characteristics such as time, location, depth, and magnitude. They are received and processed in real time to estimate arrival times of the various seismic phases, and seismic amplitudes of surface waves at the detector sites.
In section~\ref{sec:algorithm}, we describe the algorithm.
In section~\ref{sec:performance}, we describe the performance of the algorithm on the most recent gravitational-wave detector data.
In section~\ref{sec:conclusions}, we offer concluding remarks and suggest directions for future research.

\section{Algorithm}
\label{sec:algorithm}

Figure~\ref{fig:flowchart} shows the flowchart for the \emph{Seismon} pipeline, developed to mitigate the effects of teleseismic events on ground-based interferometric gravitational wave detectors. It uses event notices received from the USGS and makes time of arrival and amplitude predictions for earthquake seismic phases at sites of current detectors. Using a combination of earthquake magnitude, distance, and depth information, a prediction of the likelihood of the earthquake causing data disruption at the sites is made.

\tikzstyle{decision} = [diamond, draw, fill=blue!20,
    text width=4.5em, text badly centered, node distance=3cm, inner sep=0pt]
\tikzstyle{block} = [rectangle, draw, fill=blue!20,
    text width=5em, text centered, rounded corners, minimum height=3em]
\tikzstyle{line} = [draw, -latex']
\tikzstyle{cloud} = [draw, ellipse,fill=red!20, node distance=3cm,
    minimum height=2em]
\tikzstyle{emptyblock} = [rectangle, minimum height=3em]

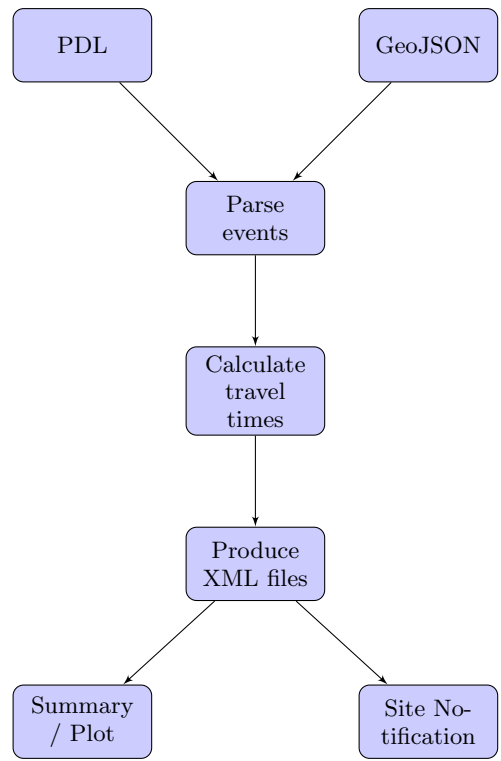
\begin{figure}[t]
 \begin{center}
 \begin{tikzpicture}[node distance = 2.3cm, auto]
    \node [emptyblock] (init) {};
    \node [block, left of=init] (PDL) {PDL};
    \node [block, right of=init] (GeoJSON) {GeoJSON};
    \node [block, below of=init] (Parse) {Parse events};
    \node [block, below of=Parse] (Traveltimes) {Calculate travel times};
    \node [block, below of=Traveltimes] (XML) {Produce XML files};
    \node [block, below of=PDL, node distance=9cm] (Plot) {Summary / Plot};
    \node [block, below of=GeoJSON, node distance=9cm] (Epics) {Site Notification};
    \path [line] (PDL) -- (Parse);
    \path [line] (GeoJSON) -- (Parse);
    \path [line] (Parse) -- (Traveltimes);
    \path [line] (Traveltimes) -- (XML);
    \path [line] (XML) -- (Plot);
    \path [line] (XML) -- (Epics);
 \end{tikzpicture}
 \end{center}
 \caption{A flow chart of the \emph{Seismon} pipeline. The USGS's Product Distribution Layer (PDL) and public GeoJSON earthquake files provide information used by \emph{Seismon} to compute estimated site arrival times and Rayleigh wave velocities.}
 \label{fig:flowchart}
\end{figure}

\subsection{Notices}

\begin{figure}[t]
\hspace*{-0.5cm}
\centering
\includegraphics[width=4in]{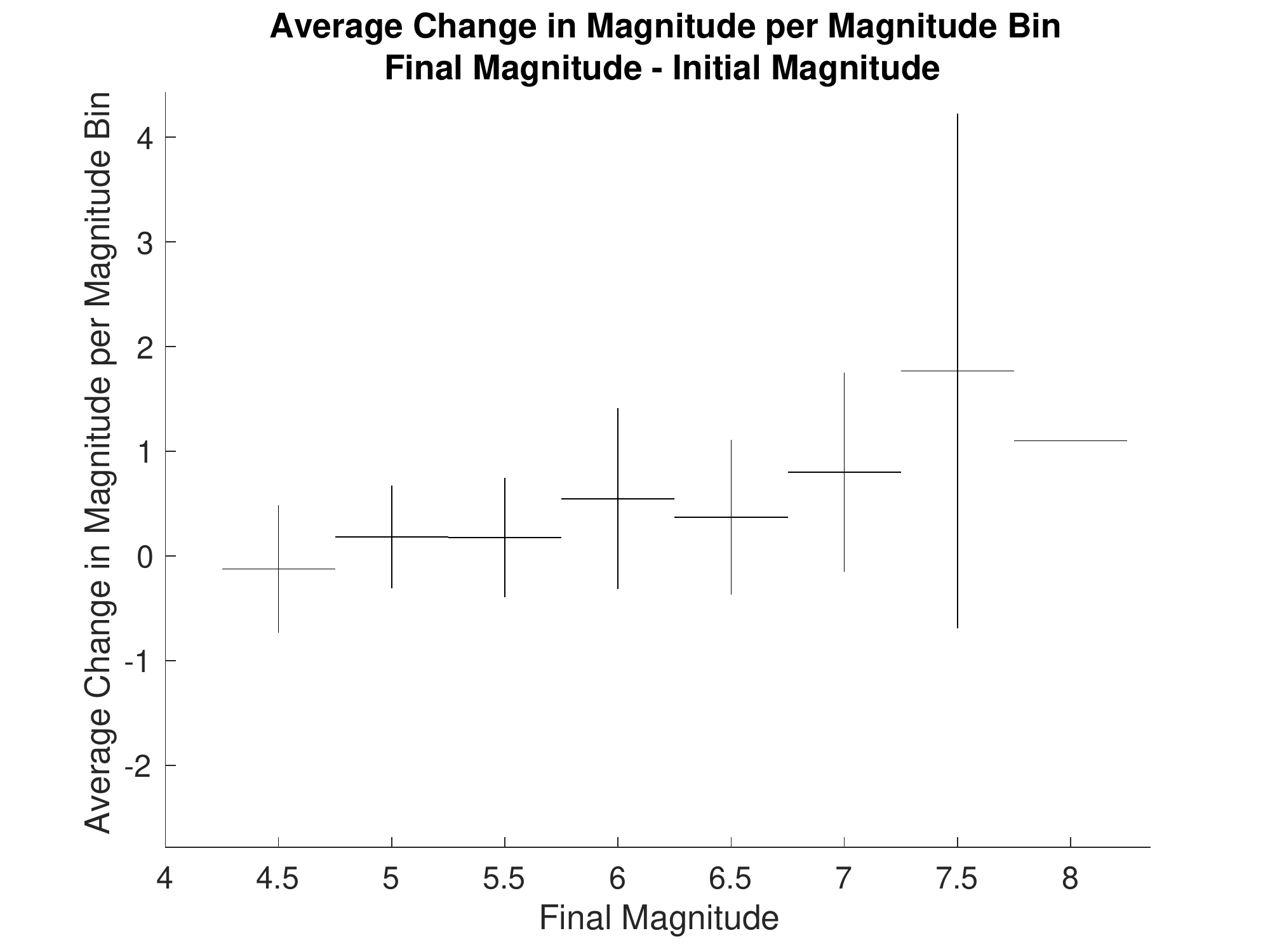}
\caption{Comparison of initial versus final magnitude estimates. Due to the desire for rapid notifications and the physical limitation of the propagation speed of seismic waves, initial magnitude estimates are distributed before all seismic stations have reported. Initial magnitudes are also calculated using rapid magnitude estimation methods. Follow-up magnitudes are calculated using more data and more time-intensive processing techniques. For the earthquakes studied here, we find magnitudes of strong earthquakes are typically underestimated in the initial analysis and estimates of magnitudes for small earthquakes are less biased.}
 \label{fig:initialfinal}
 \end{figure}

\emph{Seismon} relies on rapid earthquake notifications generated using worldwide networks of seismometers. Seismic monitoring agencies automatically detect earthquakes when P-wave arrivals are measured at several seismic stations. Preliminary estimates of the earthquake's hypocenter (latitude, longitude and depth) are refined either manually or automatically as the waves propagate to more distant stations. Initial earthquake magnitudes are based on amplitude measurements or integration of the seismic P-wave arrivals. Subsequent, more time consuming, magnitude estimation techniques use comparisons between calculated and observed seismic waveforms. These techniques also provide information about the fault orientation and slip direction.

Initial estimates of magnitude and hypocenter are distributed before the seismic waves have propagated around the Earth, and then estimates using more data and more time intensive techniques are performed. Figure~ \ref{fig:initialfinal} shows that the preliminary magnitude estimates often underestimate the final magnitude for moderate and large earthquakes. These data are from all earthquakes from the last LIGO science run, which ran from September 2015 to January 2016, comparing the earliest magnitude estimates (as would be used by \emph{Seismon}) to the final cataloged estimates computed days to weeks later.

For distant epicenters, greater warning times are possible for secondary or S-waves and surface waves than P-waves. P-waves travel at about twice the speed of S-waves, and
surface waves are delayed further since they primarily propagate in the crust and uppermost mantle where seismic velocities are slower.  Surface and S-phases generally have a stronger effect on gravitational-wave detectors due to their higher amplitudes.
Furthermore, it is likely that polarization-dependent effects also influence the impact of a seismic phase on the detector. Teleseismic P-waves produce predominantly vertical motion at the detector sites whereas S-waves and surface Love waves produce predominantly horizontal motion. Rayleigh surface waves have vertical and horizontal ground motion.

The USGS provides a number of channels for information about earthquakes on different time-scales. The earliest solutions provide hypocenter and magnitude estimates, within 5 to 20 minutes depending on its magnitude and location. At later times, moment tensor solutions and finite fault models are calculated from more data, usually arriving tens of minutes to hours after the initial notice. These solutions are distributed through the USGS's Product Distribution Layer (PDL), which has been configured to receive all notifications of earthquakes worldwide.

Several seismic networks and monitoring agencies submit earthquake alerts through this service ensuring the \emph{Seismon} pipeline will receive the most relevant notifications. The earthquake notification messages are in the form of QuakeML (XML) files \cite{ScEu2011}, although the distribution also provides image files and other related content depending on the particular earthquake. Each network that detects an earthquake provides time-tagged versions of their products, which will allow us to estimate the delay induced by the process of earthquake identification and product distribution. PDL is a cross-platform, Java-based code that runs constantly on a dedicated machine.

\subsection{Analysis}
\label{subsec:analysis}

The second step of the process is to convert event notifications to information about the time of arrival and amplitudes at the sites.
In summary, we use the location and magnitude estimates of the PDL client for two purposes. 
The first is the time of seismic wave arrivals at the gravitational-wave detectors.
The second is the ground motion at the gravitational-wave detectors.
Accurate prediction of the ground velocity amplitude based on earthquake magnitude and distance will be required to limit the false alarms. 
This equation should account for physical effects with variable parameters used to fit to the seismic data currently available.

P- and S-wave arrivals can be accurately determined given latitude, longitude, and depth information by calculating travel times using the iaspei-tau package \cite{Snoke2009} wrapped by Obspy. While surface waves experience significant dispersion, we will approximate surface waves as having a constant 3.5\,km/s speed value in our analysis below, using the associated time to compute warning times. We also provide upper and lower bound estimates of 2 and 5\,km/s. When using the data to find the peak amplitude below, we will take the beginning of the earthquake to be the P-wave arrival from iaspei-tau and the end of the earthquake to be the time associated with 2\,km/s arrivals.

The second step is to make amplitude predictions for each site. We estimate the peak amplitude of the surface waves, $\rm Rf_{amp}$, at the sites using equation (\ref{eq:Rfamp}), which we describe below. Because we have found no instances of P-wave arrivals causing the detector to lose lock, and very rare cases of the S-wave arrivals doing so, we have found it sufficient to concentrate on surface waves. This was developed as a fit to historical earthquakes at the gravitational-wave detectors. 
Eventually, it would be appropriate to determine what observational quantity is best suited to lockloss for the detectors, but for now we adopt peak amplitude due to its relative simplicity. Both the time-of-arrival and amplitude are predicted as a function of distance. This allows users of the algorithm to interpolate these metrics for their locations of interest. In general, we generate the predictions for all currently operating gravitational-wave detectors.

We now examine the historical earthquake record and predict the likely ground motion seen. We then use seismic data from onsite observations to predict how ground motion will affect the observatories. We have developed an equation attempting to account for physical effects with variable parameters used to fit to the data. Coupling strength of a source at a certain depth to Rayleigh waves, geometric amplitude evolution, and frequency-dependent scaling of the magnitude into ground displacement are taken into account. We estimate the amplitude of the surface waves, $\rm Rf_{amp}$, at the sites using the equation
\begin{equation}
{\rm Rf_{amp}} = M \frac{\rm a}{f_c^{b}} \frac{e^{-2 \pi h f_{\rm c}/c}}{r^{d}}
\label{eq:Rfamp} 
\end{equation}
where $f_{\rm c} = 10^{2.3-M/2}\,\rm$ is the corner frequency of the earthquake,  $M$ is the magnitude of the earthquake, $h$ is the depth, $r$ is the distance to the detectors, and $c$ is the speed of the surface-waves, all in SI units. Another distance-dependent exponential damping term was included initially, but it did not lead to any improvement in the amplitude prediction. One might note here that in the case of the time-of-arrival predictions we assume a constant velocity, and in the amplitude predictions we allow the value to vary. We have found this important for accounting for site specific effects as to the response to earthquakes, whereas in the time-of-arrival case, we desire an approximate time by which any changes must be made.
The difference between the prediction ${\rm Rf_{amp}}$ and the set of historical data is then minimized using the parameters $a,b,c$ and $d$.
To do so, we use a Metropolis Hastings Multi-Chain, Monte Carlo algorithm implementing adaptive simulated annealing, which statistically guarantees obtaining solutions close to global minima \cite{KiGe1983,In2000}. This algorithm was recently used in the optimization of seismometer arrays for gravity gradient noise cancellation in gravitational-wave detectors and a thorough explanation can be found in Coughlin et al. \cite{CoMu2016}. 

\subsection{Site Notification}
The final step of the process is to use the site amplitude and time-of-arrival predictions to create warnings (and possibly detector state changes) for the detectors. The algorithm analyzes the recent notifications and places a threshold on the predictions. We provide a set of site variables that contains the following information. The first is the amplitude prediction for any earthquake expected to be present.
The second is the probability of lockloss, which is discussed in the next section. The third is when this earthquake is expected to arrive at the site.
		
\section{Performance}
\label{sec:performance}

In this section, we provide a number of metrics by which we analyze the performance of \emph{Seismon}. The data we use are as follows. For LHO and LLO, the data were taken from November 2005 to October 2007 (Science Run 5 abbreviated as S5) and July 2009 to October 2010 (Science Run 6 abbreviated as S6). We will validate these fits against the latest LIGO science run (Observation Run 1 abbreviated as O1), which ran from September 2015 to January 2016. For Virgo, the data were taken from June to September 2011 (Virgo Science Run 4 abbreviated as VSR4). For GEO\,600, the data were taken from July 2010 to June 2011 (GEO High Frequency abbreviated as GEOHF). We include all earthquakes magnitude 5.0 and above with measured peak ground velocities that exceed $1\,\mu$m/s.

\subsection{Notification latency}

\begin{figure*}[t]
\hspace*{-0.5cm}
 \includegraphics[width=3.5in]{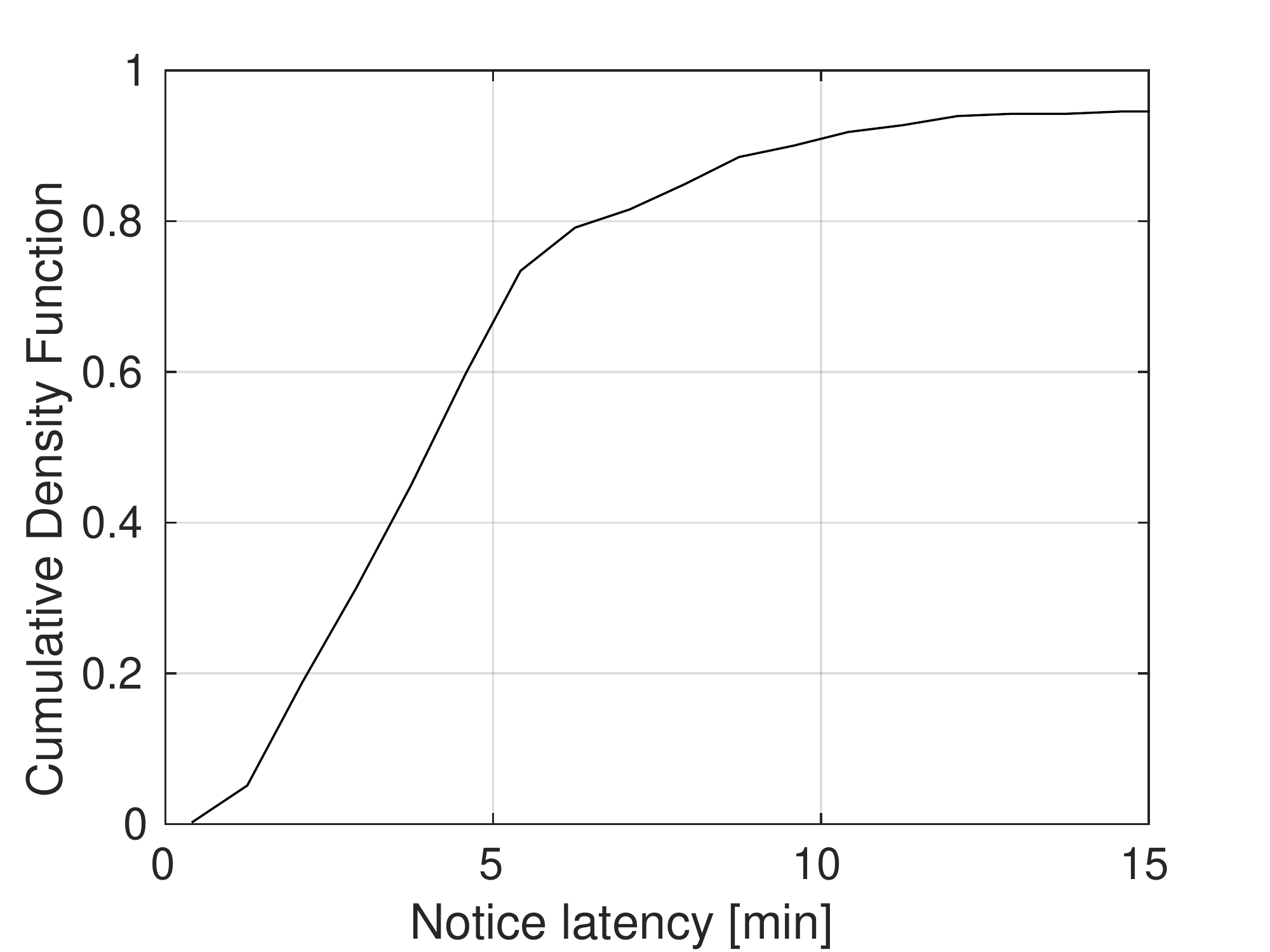}
 \includegraphics[width=3.5in]{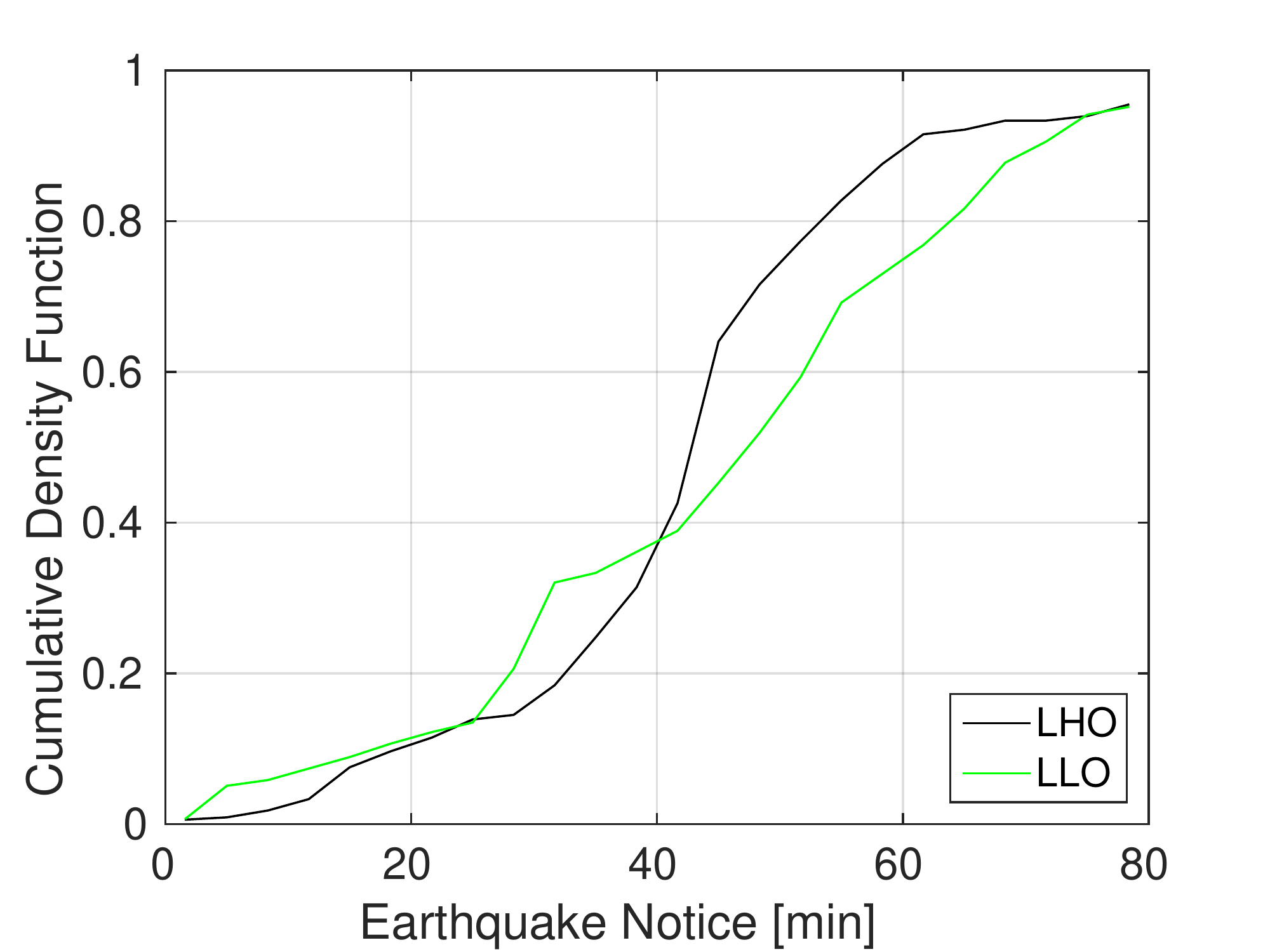}
 \caption{On the left is the time delay between the initiation of fault rupture and generation of the PDL client notification. On the right is the time delay between the earthquake notification from the PDL client and approximate arrival of surface waves at the LIGO Hanford site for global earthquakes. A majority of the earthquake locations allow for more than 10 minutes of time between notification and site arrival.}
 \label{fig:delays}
\end{figure*}

One of the most important qualities of an earthquake monitor is the notification latency, or the amount of warning time a detector has to respond to incoming seismic waves. On the left of figure~\ref{fig:delays}, we show the time delay cumulative density function between the earthquake and generation of the PDL client notification. In general, notices are generated within 5 minutes of the earthquake. On the right of figure \ref{fig:delays}, we show the cumulative probability distribution of time delays between the notification from the PDL client and approximate arrival of surface waves, assuming surface wave velocities of 3.5\,km/s. In general, there is more than 10 minutes available between notification and surface-wave arrivals. This is more than sufficient time for gravitational-wave detectors to respond by changing control configurations. For some earthquakes, this notification latency is too long. For example, for the LIGO Hanford detectors, there have been cases of earthquakes near northern California or the southern tip of Alaska that were not caught in time.

\begin{figure}[t]
\hspace*{-0.5cm}
 \includegraphics[width=3.5in]{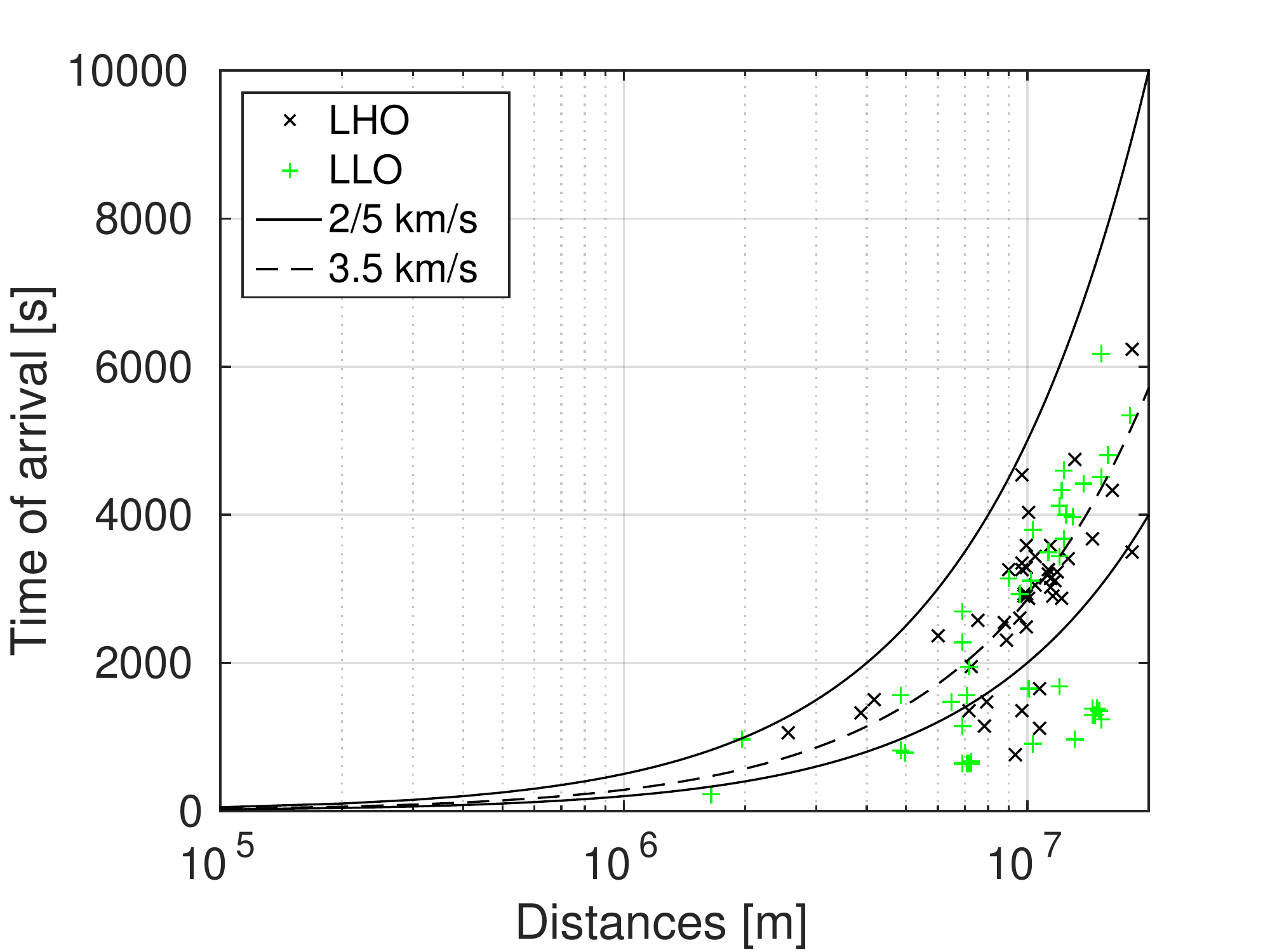}
 \caption{Time of arrival comparison for measured LHO and LLO peak velocities to constant velocity models. A majority of the earthquake arrivals fall within the 2-5\,km/s bounds, predominantly clustered around 3.5\,km/s in the middle. Arrivals outside of these bounds likely are due to either P- and S-wave arrivals or misidentification due to overlapping earthquakes.}
 \label{fig:TOA}
\end{figure}

Figure~\ref{fig:TOA} shows the time-of-arrivals for the peak ground velocity measured at LHO and LLO, as well as 2\,km/s, 3.5\,km/s, and 5\,km/s bounds. In this paper, we adopt 3.5\,km/s as the nominal surface-wave arrivals for computing potential warning times. \emph{Seismon} provides 2\,km/s, 3.5\,km/s, and 5\,km/s arrival estimates to cover the potential surface arrivals. There are a handful of instances where the maximum occurred outside of this interval, and this is likely due to either P- and S- wave arrivals or misidentification due to overlapping earthquakes.

\subsection{Ground Velocity Prediction Performance}

\begin{figure*}[t]
\hspace*{-0.5cm}
 \includegraphics[width=3.5in,trim = 2.5cm 1.5cm 2.5cm 1.5cm, clip=true]{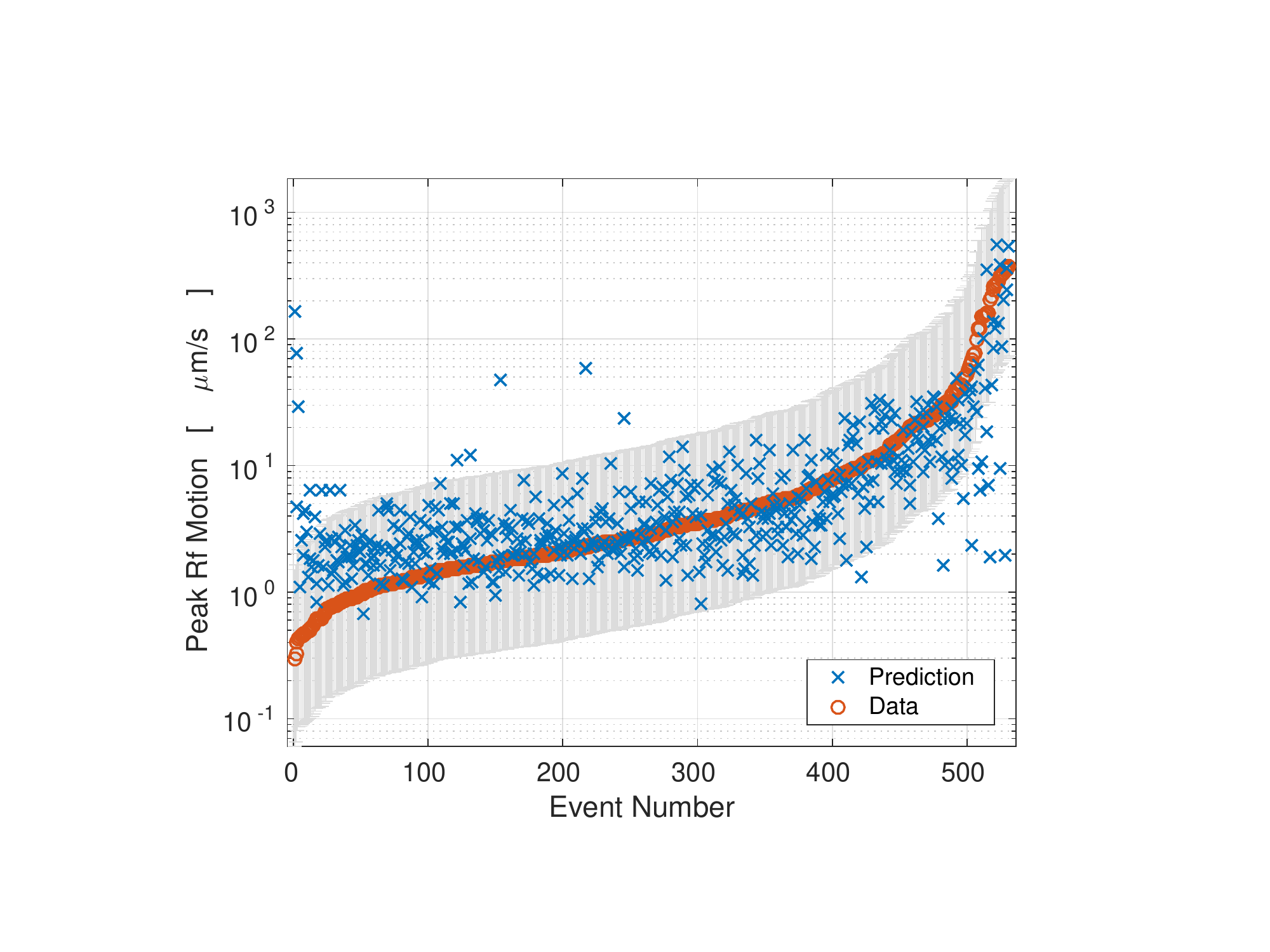}
 \includegraphics[width=3.5in,trim = 2.5cm 1.5cm 2.5cm 1.5cm, clip=true]{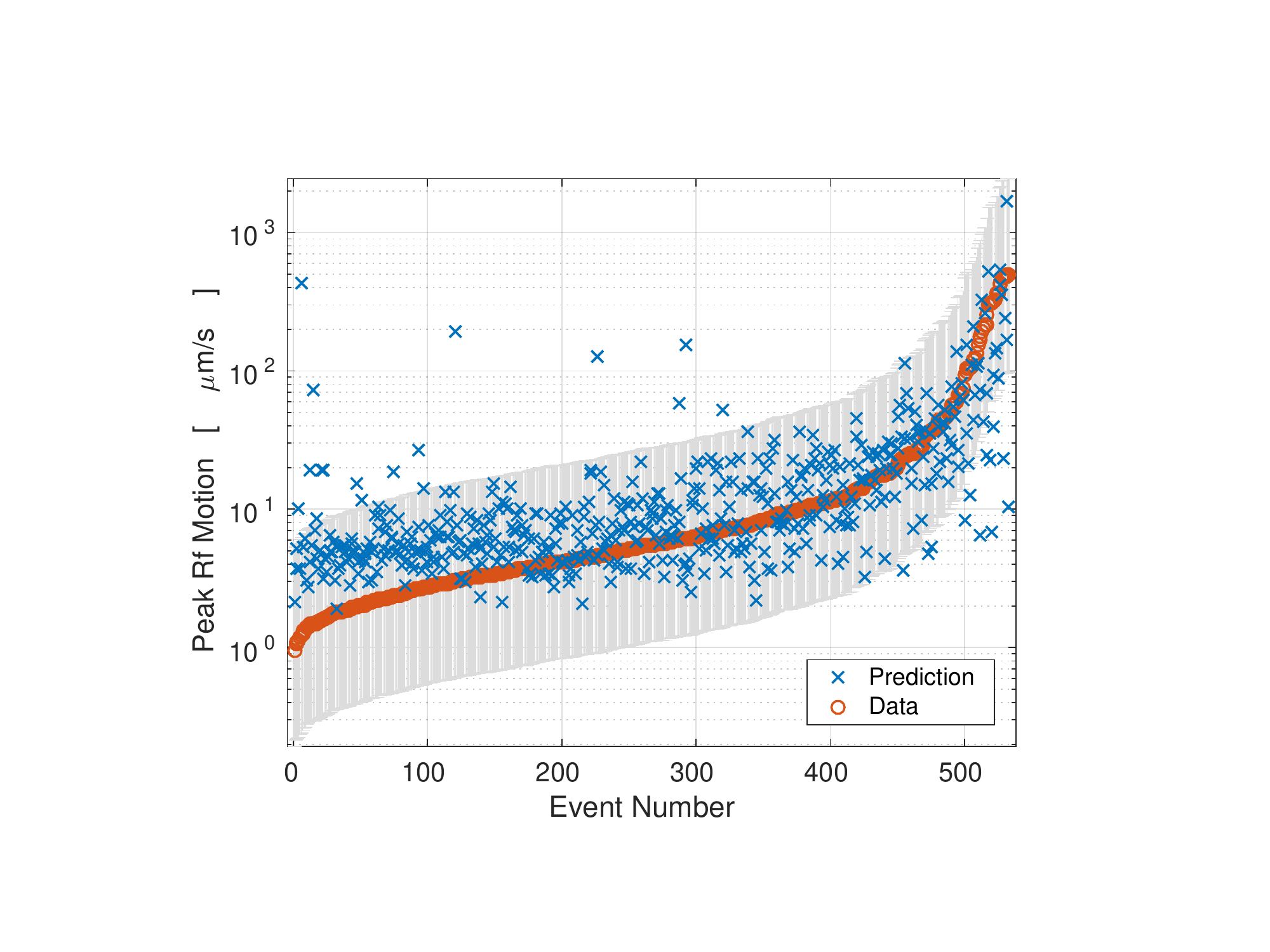}
 \caption{Fit of peak velocities seen during S5-S6 at the interferometers (LHO and LLO) to equation \ref{eq:Rfamp}.  Fit parameters are estimated from S5-S6 data using final event parameter estimates. The events have been ordered by their measured peak ground velocity (in blue) and the green crosses correspond to the prediction from the equation. About 90\% (LHO and LLO) of events are within a factor of 5 of the predicted value.}
 \label{fig:regression}
\end{figure*}

\begin{figure}[t]
\hspace*{-0.5cm}
 \includegraphics[width=3.5in]{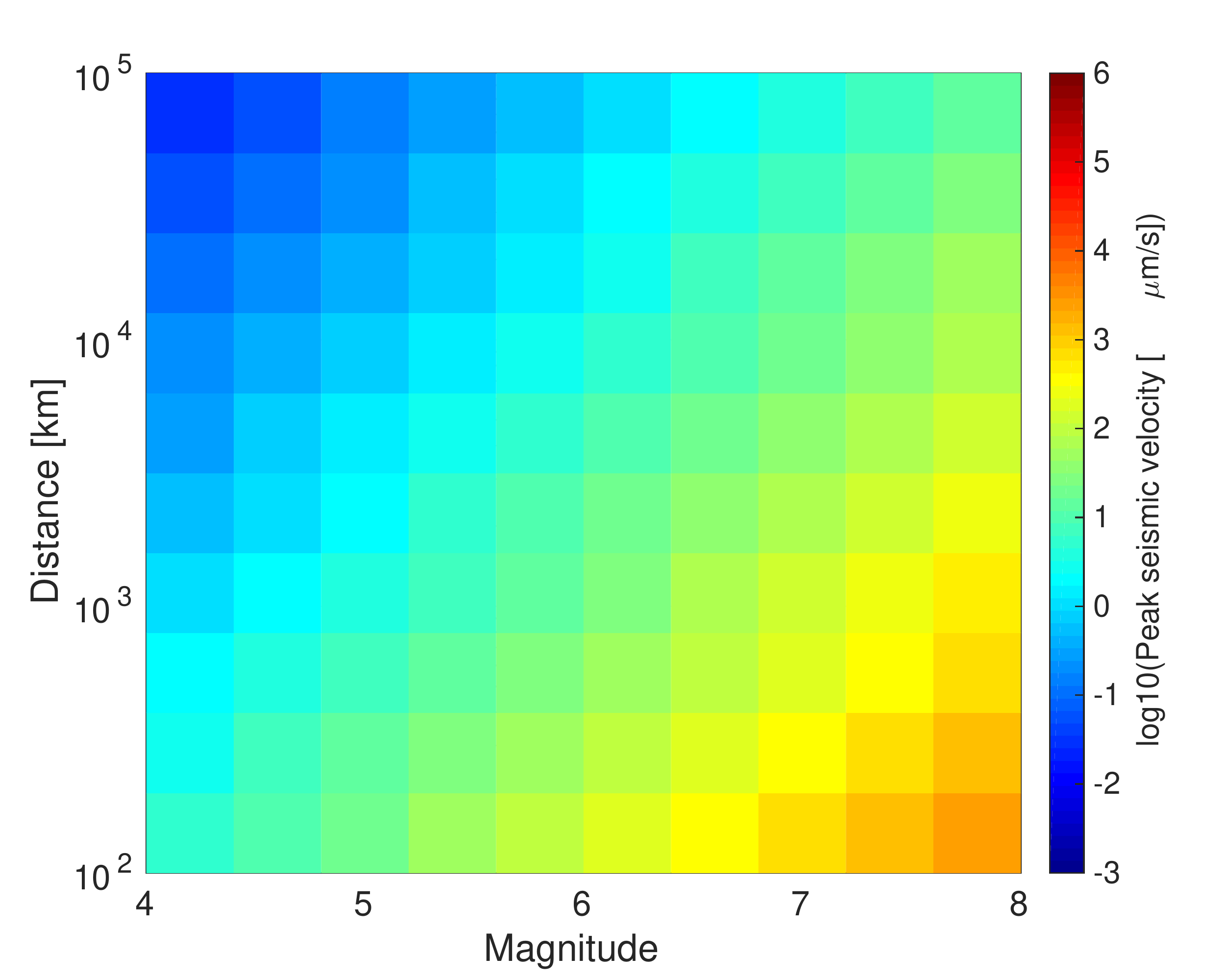}
 \caption{The predicted peak ground velocity as a function of magnitude and distance for LHO (LLO is similar).}
 \label{fig:MvsR}
\end{figure}

As described in section~\ref{subsec:analysis}, we fit free parameters in equation~\ref{eq:Rfamp} to measured peak surface-wave velocities at current gravitational-wave detectors. Best-fit parameter values are summarized for the gravitational-wave detectors in this study in table~\ref{table:fit} assuming that all physical parameters are in SI units. The regression is shown in figure \ref{fig:regression} for both LHO and LLO gravitational-wave interferometers.
There is significant scatter in the parameter values across the detectors. In particular, the surface wave speed parameter varies about an order of magnitude between GEO and the other detectors. This is due to the significant degeneracy between parameters in the adopted model and that many parameter combinations give similar results. It is possible that in the future, an equation with fewer degenerate parameters could be found to alleviate this issue.
Figure~\ref{fig:MvsR} shows the peak ground velocity as a function of magnitude and distance for the models. Based on the above equations, we expect that earthquakes with magnitudes greater than 5 can exceed ground velocities of $1\,\mu$m/s. Gravitational-wave detector operators have found that $1\,\mu$m/s is the approximate threshold during which detectors can continue to take data, and so it is useful to have a quick visual as to expected ground velocities for any given earthquake.
\begin{table}[]
\centering
\begin{tabular}{|c|c|c|c|c|}
\hline
Detector & $a$ & $b$ & $c$ & $d$ \\ \hline
LHO & 0.16 & 1.31 & 4672.83 & 0.83 \\ \hline
LLO & 0.16 & 1.31 & 4672.83 & 0.81 \\ \hline
VIRGO & 1.60 & 0.89 & 4992.70 & 0.83 \\ \hline
GEO & 8.65 & 1.92 & 324.52 & 1.40 \\ \hline
\end{tabular}
\caption{Best-fit parameters to the peak velocities seen at the interferometers to equation~\ref{eq:Rfamp}.}
\label{table:fit}
\end{table}

\begin{figure}[t]
\hspace*{-0.5cm}
 \includegraphics[width=3.5in]{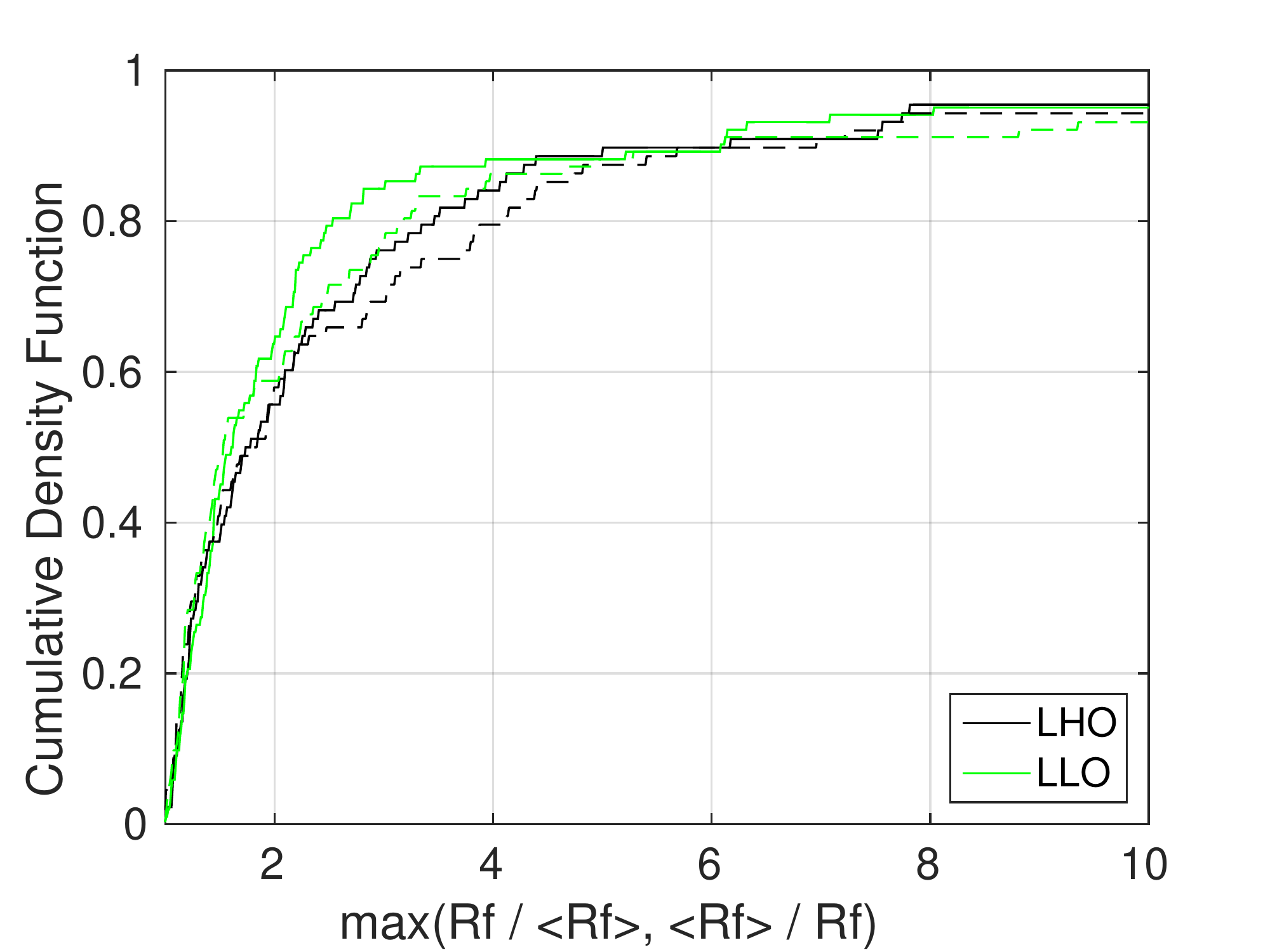}
 \caption{Performance of estimation of peak velocities seen during O1 at the interferometers (LHO and LLO) using fit parameters estimated from S5-S6 data. The x-axis gives the maximum of the ratio between the estimated and measured peak ground velocities and vice versa. The solid lines use the final earthquake parameter estimates while the dashed lines use the preliminary earthquake estimates. About 90\% of events are within a factor of 5 of the predicted value. The difference in fit parameters due to use of preliminary notices is minimal.}
 \label{fig:regressionperf}
\end{figure}

Another important quality for an earthquake monitor is the accuracy of the ground-motion amplitude prediction and the time-of-arrival.
The ground-motion amplitude performance is evaluated against the most recent LIGO science run (Observing Run 1) from September 2015 to January 2016, in figure \ref{fig:regressionperf}. About 90\% of events are within a factor of 5 of the predicted value, while those that are not are almost exclusively events that are due to the overlap of many events. This occurs often during aftershocks of large earthquakes. As the largest event is the important one, these are unimportant for predictions. 

\begin{figure}[t]
\hspace*{-0.5cm}
 \includegraphics[width=3.5in]{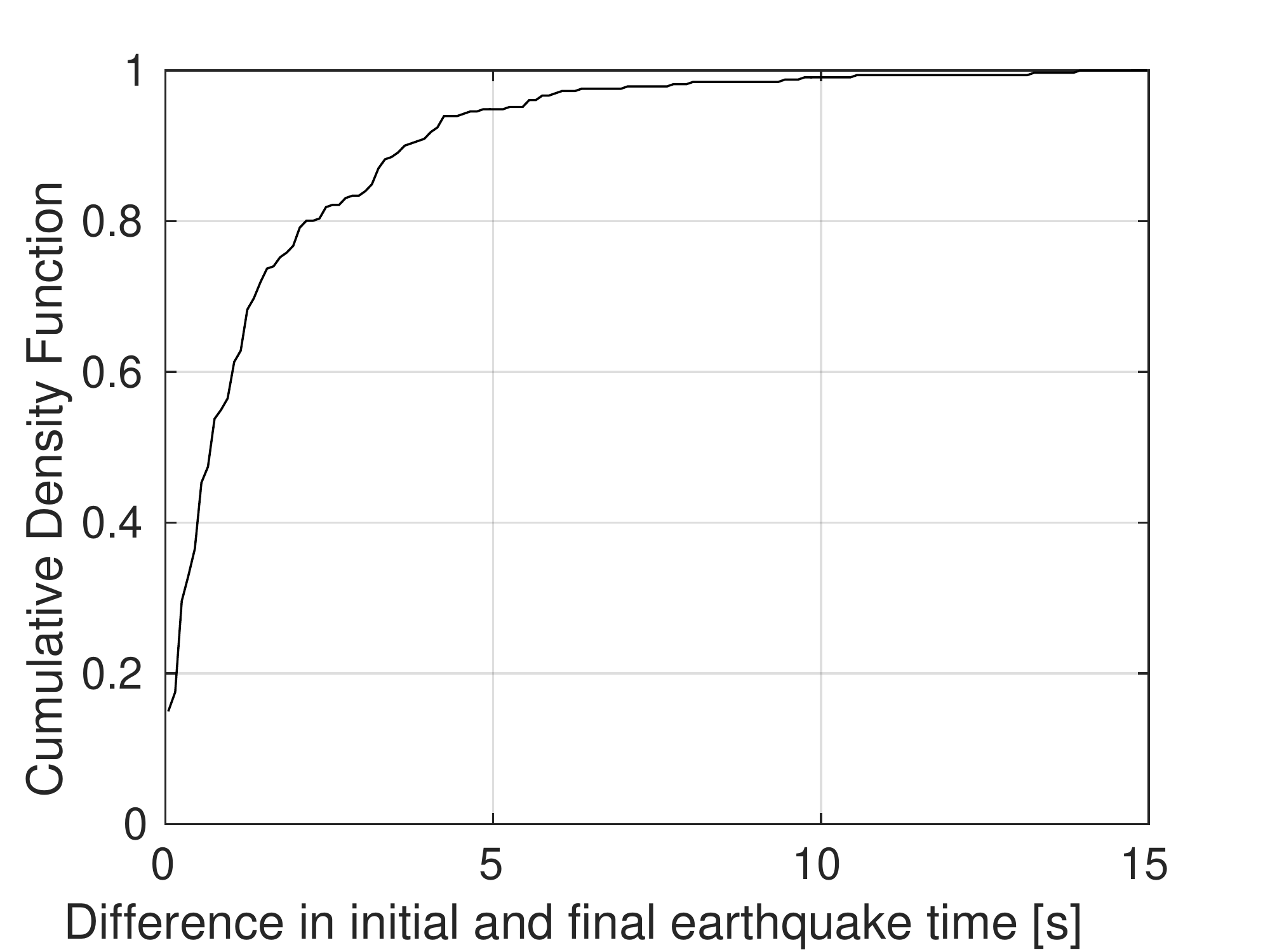}
 \caption{Difference between the initial and final estimates of the earthquake time. About 90\% of early estimates are within 4\,s of the final time.}
 \label{fig:initialvsfinal}
\end{figure}

As mentioned above, \emph{Seismon} uses the earliest available notices for making time-of-arrival and amplitude predictions. Because the earliest notices may only rely on a few seismometers, as well as the fact that large earthquakes do not fault all at once, the estimates for both magnitude, depth, location, and time can be off. In figure \ref{fig:regressionperf}, we show the difference of predicted peak velocities seen during O1 at the interferometers (LHO and LLO) using the initial and final estimates. This is a smaller error than from the regression. 
In figure~\ref{fig:initialvsfinal}, we show the difference between the initial and final estimates of the earthquake time. About 90\% of early estimates are within 3\,s of the final time, which is much smaller than the latency from the generation of the notice itself.
For these reasons, the use of the early notices is not a major source of systematic error for \emph{Seismon}.

\subsection{Gravitational-wave detector lockloss prediction performance}

\begin{figure*}[t]
\hspace*{-0.5cm}
 \includegraphics[width=3.5in]{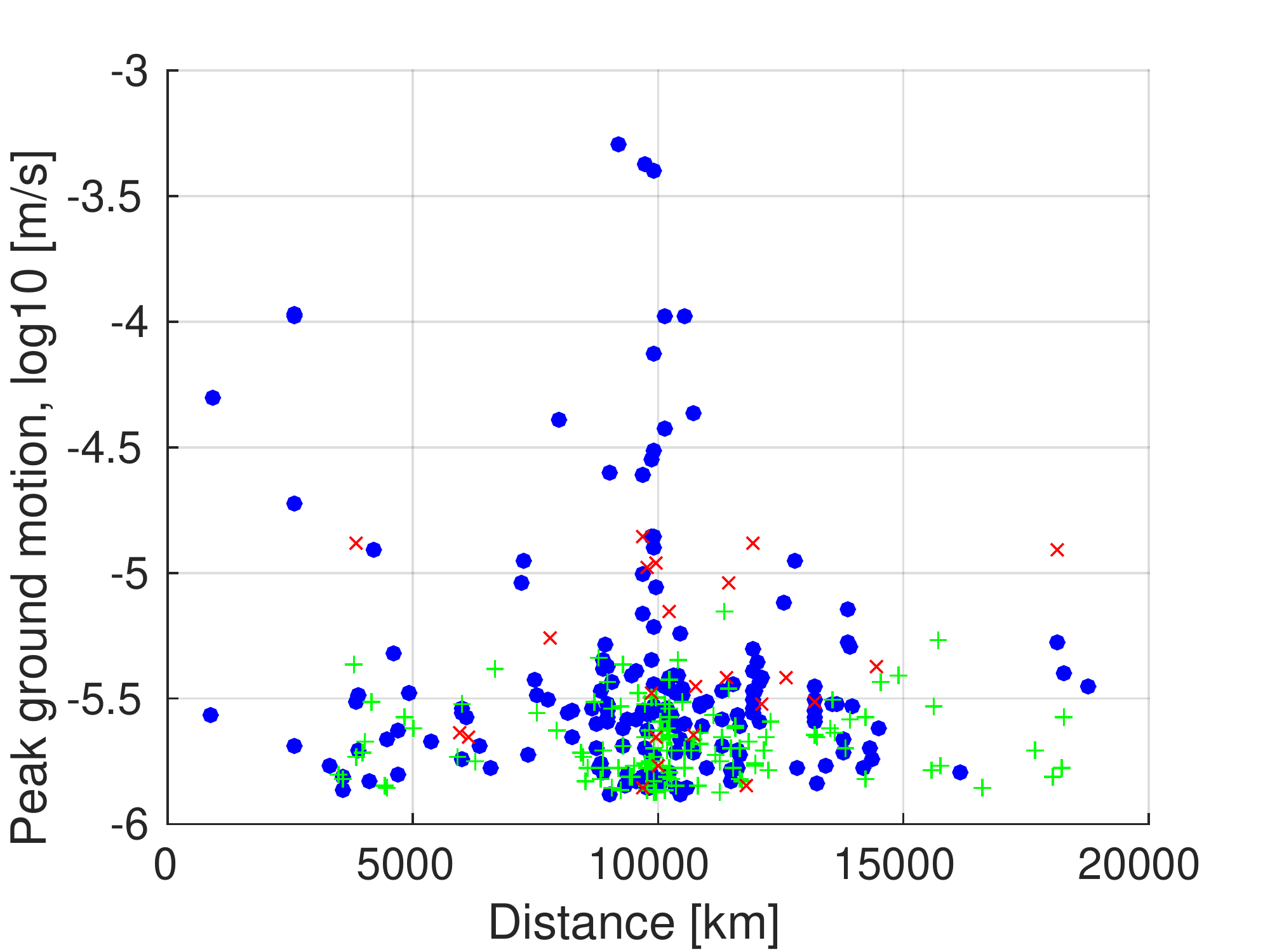}
  \includegraphics[width=3.5in]{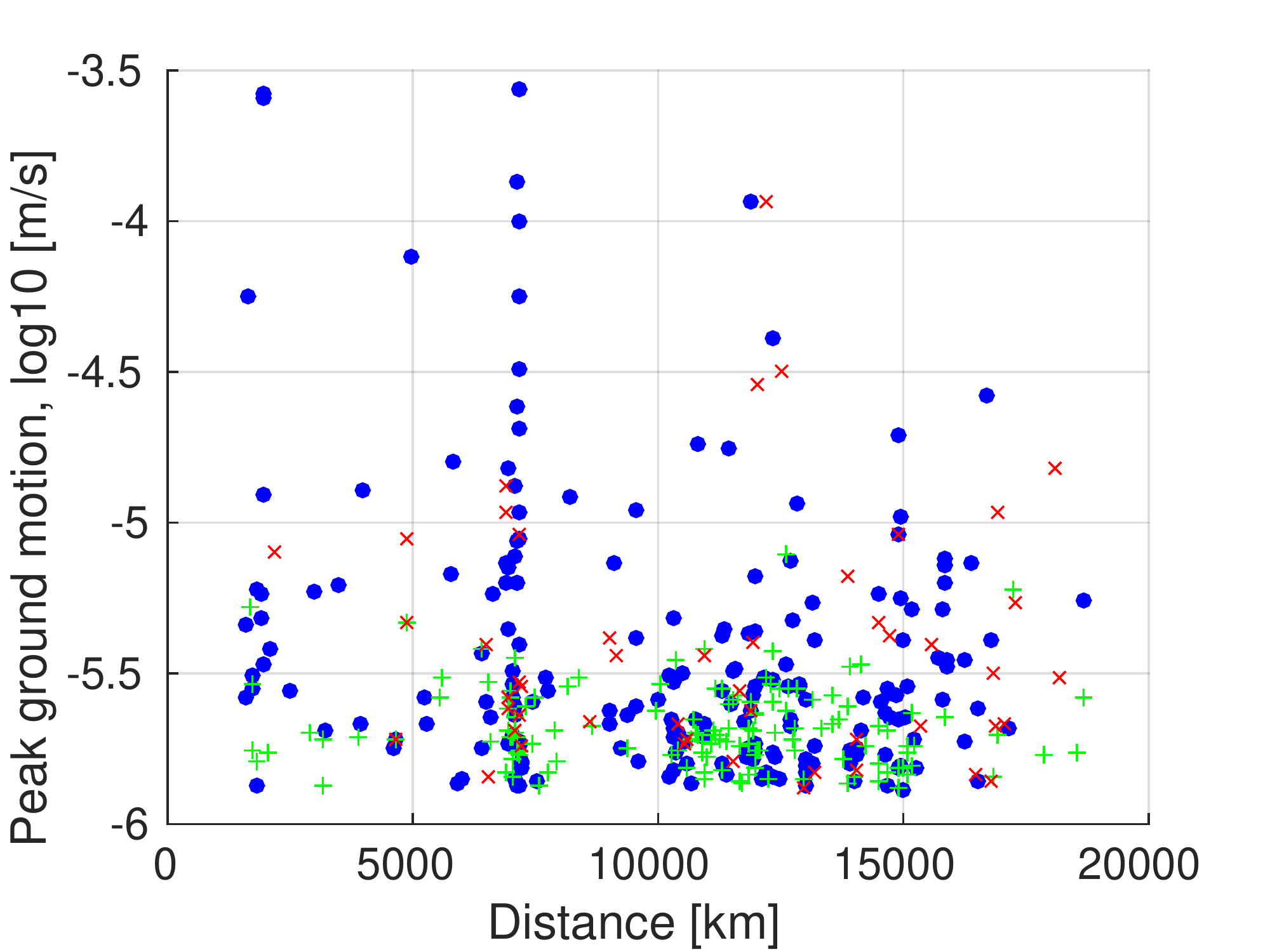}
 \caption{Lockloss as a function of predicted peak velocity vs. earthquake distance for the gravitational-wave detectors. LHO is on the left and LLO is on the right. Blue circles correspond to times when the detector was not locked, red crosses when the detector lost lock, and green plus signs when the detector stayed locked.}
 \label{fig:lockloss}
\end{figure*}

An earthquake monitor will only be useful for gravitational-wave detectors if it can be determined which earthquakes cause the loss of data and which will not affect the detector in a significant way.
We now measure the amplitude of the seismic ground motion that causes the detector to lose lock. To do so, we take all known earthquakes above magnitude 5.0 and compute their arrival times. 
We also determine the times that the gravitational-wave detectors fell out of lock during these times. 
Figure~\ref{fig:lockloss} shows these times, both for those times when locklosses occurred, when they did not, and when the detector was not locked as a function of peak ground velocity. 
The plot shows that while in general ground velocities greater than about 5\,$\mu$m/s lead to lockloss, the situation is complicated at lower ground velocities. This motivates using more than ground velocity to predict lockloss, as for example the spectrum of ground motion or the direction of propagation of seismic waves with respect to the detector orientation.

It is of significant interest to determine the earthquake parameters that cause the detectors to lose lock due to the ground velocities they create.
Given that \emph{Seismon} is an early warning system, the only parameters available for use are those returned by USGS in low latency, which are magnitude, depth and location (and thus distance). 
In addition, we can use the predicted ground velocity derived from these parameters.
The goal is to predict the outcome of the interferometer lock status based on these parameters.

\begin{figure*}[t]
\hspace*{-0.5cm}
 \includegraphics[width=3.5in, trim = 2.5cm 1.5cm 2.5cm 1.5cm, clip=true]{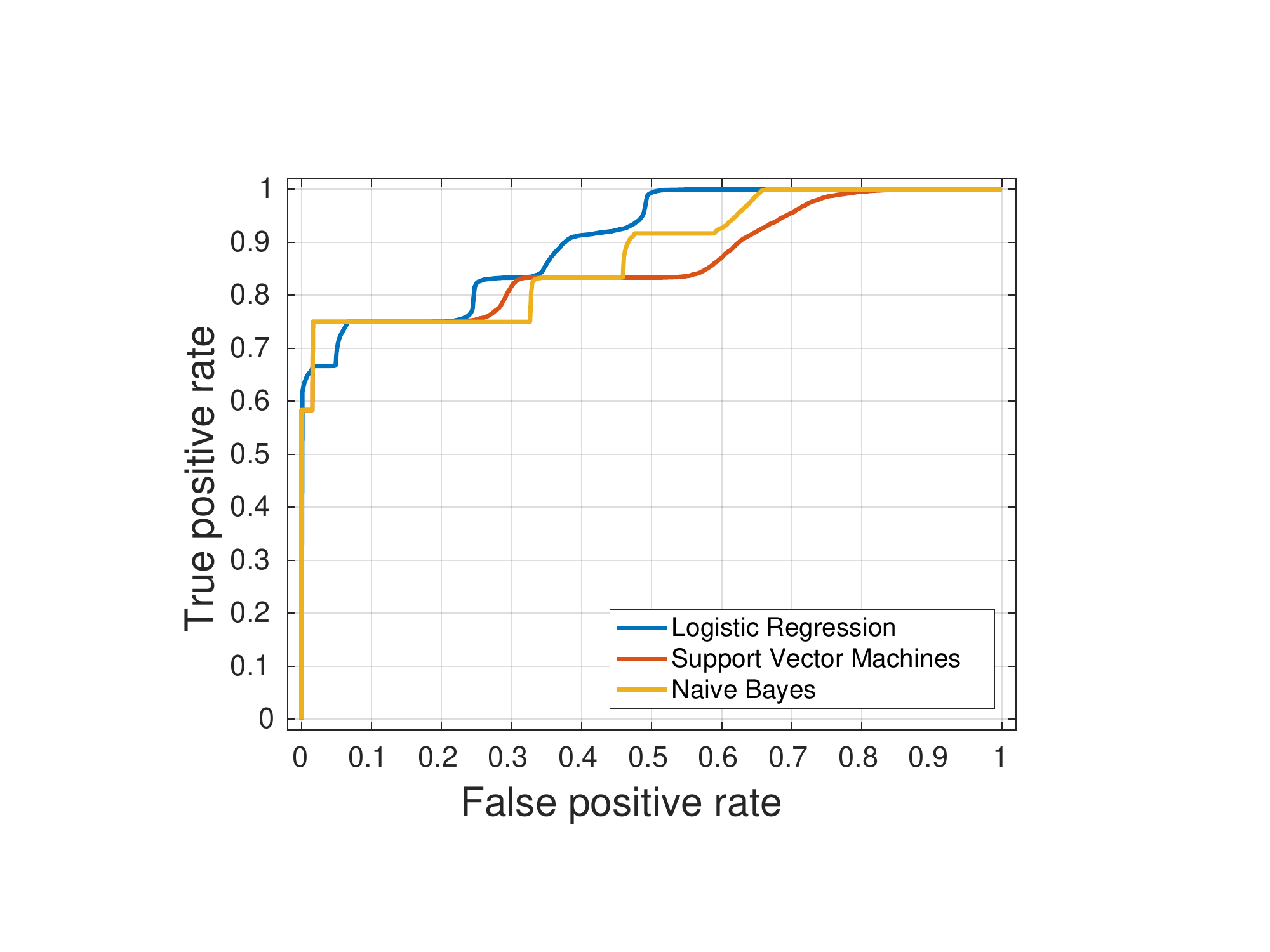}
  \includegraphics[width=3.5in, trim = 2.5cm 1.5cm 2.5cm 1.5cm, clip=true]{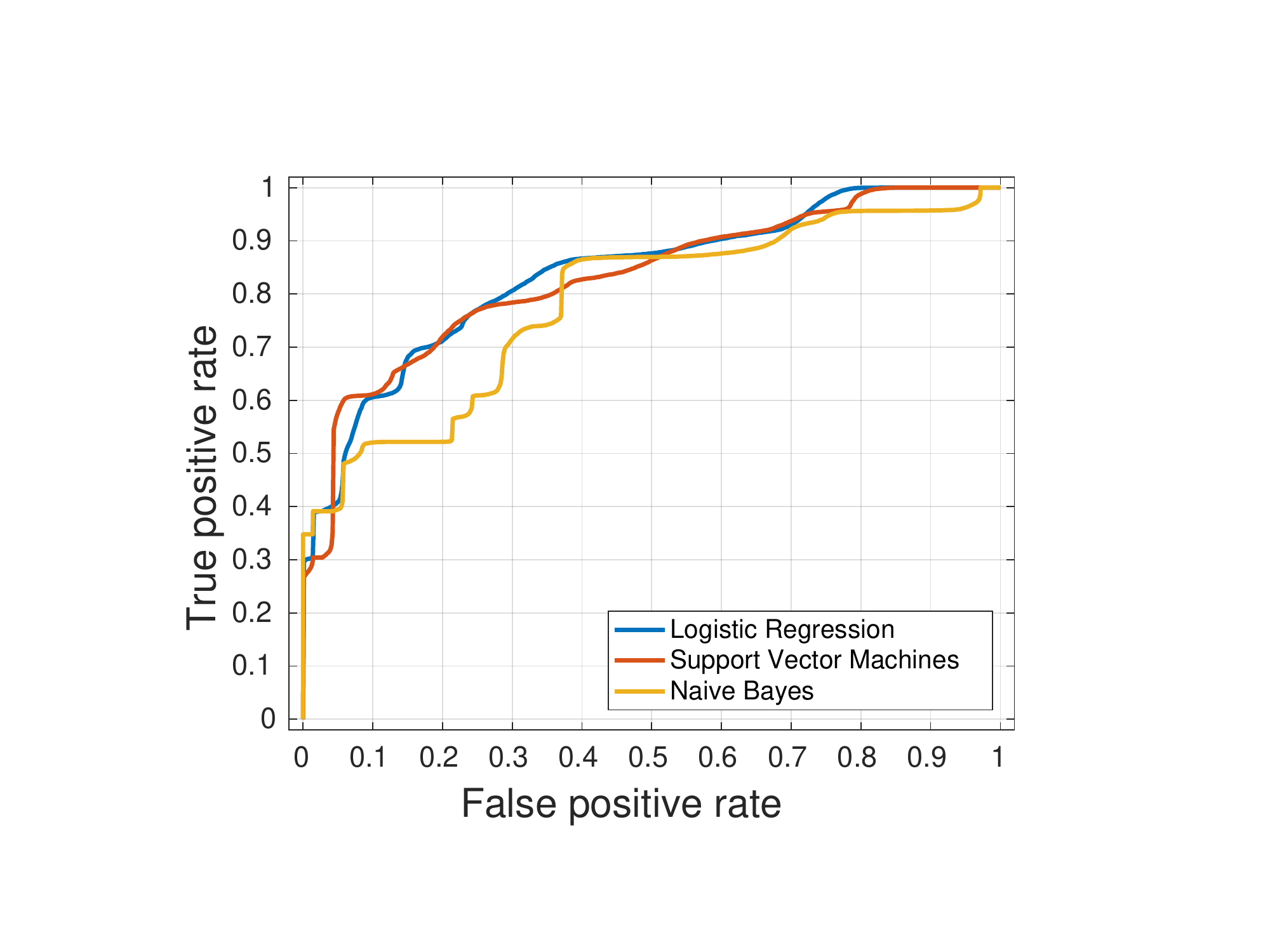}
 \caption{Performance comparison of different machine learning classifiers. LHO is on the left and LLO is on the right. True positive rate is the ratio of the sum of predicted positive condition actually being true to the sum of all actually positive conditions. Positive condition here refers to a lockloss prediction by the classifier which in general can be true or false. False positive rate is the ratio of the sum of predicted positive condition being false to the sum of all actually negative conditions. Classifier prediction about the detector being in lock forms the negative condition. A threshold value on classifier output (usually a scaled number between 0 and 1 with value close to unity indicative of lockloss) is varied to generate the receiver operator characteristic curve (ROC).
 }
 \label{fig:MLA_comparison}
\end{figure*}

In the following, we will use a machine learning algorithm to develop a lockloss prediction model. Machine learning algorithms, which are useful for classifying and predicting outcomes for various data analysis problems, have been used in the past with great success in gravitational-wave data analysis \cite{BiBl2013,KyHa2015}. We first compare the performance of different machine learning algorithms aimed at modelling lockloss prediction from the given input parameters.  All three classifiers, Logistic Regression \cite{mccullagh_glm}, Naive Bayes \cite{John_NaiveBayes}, and Support Vector Machine \cite{Burges_SVM}, yield comparable performance with logistic regression giving the best result as is evident from the receiver operator characteristic curve (ROC) shown in Figure~\ref{fig:MLA_comparison}. 
The classifier with the maximal area under the curve is usually chosen over the others. 

\emph{Seismon} makes lockless predictions using the threshold value obtained from optimal operating point of the ROC curve. 	
The idea is that by setting the false-alarm rate to a certain threshold, we can find an optimum point of efficiency for predicting the outcome. In the analysis that follows, we use $2/3$ of the earthquakes for training and  $1/3$ of the earthquakes for the testing set. 
In general, there is a trade-off between false-alarm probability and efficiency standard probability. The more false alarms one is willing to accept, the higher the rate of earthquakes that will result in lockloss will be caught. For example, if we adopt a false-alarm probability threshold of 0.5, between 90 -- 100\% of earthquakes can be caught. While there is potential frustration of using a detector configuration that is, by design, more noisy, during false positive events, as the switch between detector configuration states is orders of magnitude faster than lock acquisition for gravitational-wave detectors, it is worth the trade-off. This is of course not to say that potential improvements are not important, and we have begun to explore potential direction dependent effects to improve the statistics.

\section{Conclusion}
\label{sec:conclusions}

In this paper, we have discussed the problem of earthquakes for gravitational-wave detectors and a pipeline designed to minimize their impact. 
We characterize this pipeline in terms of the warning time for these experiments.
We have shown that the earthquake warning system can both predict likely earthquake arrival times and ground velocity amplitudes. 

A code that performs these steps is available at https://github.com/ligovirgo/seismon/ for public download. Hopefully, this will allow other researchers to easily use the fits. Required inputs are the latitude and longitude of the site and magnitude, latitude, longitude and depth of the source.

In the future, the \emph{Seismon} algorithm will be put in operation in the next science run. It will require coordination between the low-latency notification software and the detector control systems to maximize the utility of the system. 
Further effort needs to be spent on investigating the effect of strong ground motion on the detector control system.
This will include studies of the control configuration best for riding out times of large ground motion.

\section{Acknowledgments}
MC was supported by the National Science Foundation Graduate Research Fellowship
Program, under NSF grant number DGE 1144152. 
NM acknowledges Council for Scientific and Industrial Research (CSIR), India, for providing financial support as Senior Research Fellow.  
LIGO was constructed by the California Institute of Technology and Massachusetts Institute of Technology with funding from the National Science Foundation and operates under cooperative agreement PHY-0757058.
This paper has been assigned LIGO document number LIGO-P1600321.
The authors would like to thank Dr. Jenne Driggers for a detailed reading of an early version of the manuscript.

\bibliographystyle{unsrt}
\bibliography{references}

\end{document}